\documentclass{article}
\usepackage{latexsym}
\usepackage{epsfig}
\usepackage{amsmath}
\usepackage{amssymb}
\usepackage{amsthm}
\newtheorem{theorem}{Theorem}[section]
\newtheorem{corollary}[theorem]{Corollary}

\newtheorem{lemma}[theorem]{Lemma}

\theoremstyle{remark}
\newtheorem{remark}[theorem]{Remark}

\theoremstyle{definition}
\newtheorem{definition}[theorem]{Definition}

\DeclareMathOperator\id{id}

\DeclareMathOperator\interior{int}

\begin{document}

\title{Concurrence of Lorentz-positive maps}

\author{Roland Hildebrand \thanks{%
LMC, Universit\'e Joseph Fourier, Tour IRMA, 51 rue des
Math\'ematiques, 38400 St.\ Martin d'H\`eres, France ({\tt
roland.hildebrand@imag.fr}).}}

\maketitle

\begin{abstract}
Let ${\cal H}(d)$ be the space of complex hermitian matrices of size $d \times d$ and let $H_+(d) \subset {\cal H}(d)$
be the cone of positive semidefinite matrices. A linear operator $\Phi: {\cal H}(d_1) \to {\cal H}(d_2)$ is said to be positive
if $\Phi[H_+(d_1)] \subset H_+(d_2)$. The concurrence $C(\Phi;\cdot)$ of a positive operator $\Phi: {\cal H}(d_1) \to {\cal H}(d_2)$
is a real-valued function on the cone $H_+(d_1)$, defined as the largest convex function which coincides with
$2\sqrt{\sigma_2^{d_2}(\Phi(\xi\xi^*))}$ on all rank 1 matrices $\xi\xi^* \in H_+(d_1)$. Here $\sigma_2^d: {\cal H}(d) \to {\mathbb R}$
denotes the second symmetric function, defined by $\sigma_2^d(A) = \sum_{i < j} \mu_i\mu_j$,
where $\mu_1,\dots,\mu_d$ are the eigenvalues of $A$. The concurrence of a bipartite density matrix $X$ is defined
as the concurrence $C(\Phi;X)$ with $\Phi$ being the partial trace.

A analogous concept can be considered for Lorentz-positive maps. Let $L_n \subset {\mathbb R}^n$ be the
$n$-dimensional Lorentz cone. Then a linear map $\Upsilon: {\mathbb R}^m \to {\mathbb R}^n$ is called
Lorentz-positive if $\Upsilon[L_m] \subset L_n$. For this class of maps we are able to compute the
concurrence explicitly.

This allows us to obtain formulae for the concurrence of positive operators having ${\cal H}(2)$ as input space
and consequently of bipartite density matrices of rank 2. Namely, let $\Phi: {\cal H}(2) \to {\cal H}(d_2)$ be a positive operator,
and let $\lambda_1,\dots,\lambda_4$ be the generalized eigenvalues of the pencil $\sigma_2^{d_2}(\Phi(X)) - \lambda \det X$,
in decreasing order. Then the concurrence is given by the expression $C(\Phi;X) = 2\sqrt{\sigma_2^{d_2}(\Phi(X)) - \lambda_2 \det X}$.
As an application, we compute the concurrences of the density matrices of all graphs with 2 edges.

Similar results apply for a function which we call $I$-fidelity, with the second largest generalized eigenvalue $\lambda_2$ replaced by the smallest
generalized eigenvalue $\lambda_4$.
\end{abstract}

\section{Introduction}

The concurrence is a scalar function initially introduced to quantify the entanglement of bipartite density
matrices describing the mixed states of 2-qubits \cite{HillWootters97}. In \cite{HillWootters97}, an explicit
formula for the concurrence of $2 \otimes 2$ bipartite density matrices of rank 2 was obtained. In a
subsequent paper \cite{Wootters98}, Wootters generalized this formula to $2 \otimes 2$ bipartite density
matrices of arbitrary rank. Further generalizations were achieved by Uhlmann \cite{Uhlmann00}. He considered
real-valued functions $f(\xi\xi^*)$ on the set of pure states and introduced the convex roof of $f$, which
is the largest convex extension $f(\rho)$ to the set of all density matrices. Similarly, the concave roof is
the smallest concave extension of $f$. Uhlmann derived an explicit formula for the convex roof of the
function $f(\xi\xi^*) = |\xi^*\Theta\xi|$, where $\Theta$ is an arbitrary anti-linear hermitian operator
acting on the state vector $\xi$. He called this convex roof $\Theta$-concurrence. It then turns out that
the concurrence for $2 \otimes 2$ bipartite density matrices just equals the $\Theta$-concurrence for a
special anti-linear hermitian operator $\Theta$ acting on ${\mathbb C}^4$, and Wootters formula is a special
case of Uhlmanns formula for $\Theta$-concurrences. Uhlmann derived a similar formula for the $\Theta$-fidelity,
which he defined as the concave roof of the function $f(\xi\xi^*) = |\xi^*\Theta\xi|$.

Rungta et al.\ \cite{Rungta01} defined the $I$-concurrence of arbitrary bipartite density matrices as convex
roof of the function $f(\xi\xi^*) = 2\sqrt{\sigma_2(tr_1(\xi\xi^*))}$, where $\sigma_2$ is the second
symmetric function of a matrix and $tr_1$ is the partial trace with respect to the first of the two subsystems.
Since the $I$-concurrence is the unique natural generalization of the concurrence as defined in
\cite{HillWootters97}, we will henceforth call it simply concurrence. Rungta and Caves \cite{RungtaCaves03}
computed the concurrence explicitly for $d \otimes d$ bipartite density matrices of isotropic states, i.e.\
convex combinations of the maximally mixed state and a maximally entangled state. Osborne \cite{Osborne05}
obtained a formula for the tangle, i.e.\ the convex roof of the function $f(\xi\xi^*) =
4\sigma_2(tr_1(\xi\xi^*))$, of rank two density matrices, using essentially the main idea in
\cite{HillWootters97}. Uhlmann \cite{Uhlmann05} then went up one abstraction level and replaced the partial
trace $tr_1$ in the formula $f(\xi\xi^*) = 2\sqrt{\sigma_2(tr_1(\xi\xi^*))}$ by an arbitrary positive
operator $\Phi$, i.e.\ he defined the concurrence $C(\Phi;\rho)$ of a state $\rho$ with respect to the operator
$\Phi$ as the convex roof of the function $f(\xi\xi^*) = 2\sqrt{\sigma_2(\Phi(\xi\xi^*))}$. He then
showed that when $\Phi$ is a completely positive map of rank and length two, the concurrence $C(\Phi;\cdot)$
can be reduced to the $\Theta$-concurrence for a suitable anti-linear hermitian operator $\Theta$.

As noted above, the $\Theta$-fidelity differs from the $\Theta$-concurrence by the substitution of the convex roof
by the concave roof. By analogy, we are tempted to introduce a function called $I$-fidelity by replacing the
convex by a concave roof in the definition of the $I$-concurrence. For a bipartite density matrix,
the $I$-fidelity will then be the smallest concave extension of the function $f(\xi\xi^*) = 2\sqrt{\sigma_2(tr_1(\xi\xi^*))}$
defined on the pure states. The $I$-fidelity $F(\Phi;\cdot)$ of a positive operator $\Phi$
is then the concave roof of the function $f(\xi\xi^*) = 2\sqrt{\sigma_2(\Phi(\xi\xi^*))}$.

In the present contribution, we generalize the ideas of \cite{HillWootters97} in a completely different
direction. Our point of departure is that the second symmetric function is a quadratic form with signature
$(+--\cdots-)$. Such forms are intimately linked with the second-order cones, or Lorentz cones, which are
defined in real vector spaces ${\mathbb R}^n$ of any dimension. Like the cones of positive semidefinite
matrices, the Lorentz cones are so-called {\sl self-scaled} cones, i.e.\ cones of squares for some Jordan
algebra \cite{KoecherBraun}. The structure defined on ${\mathbb R}^n$ by the Jordan algebra permits to
introduce notions like eigenvalues, rank, trace and determinant for arbitrary elements of ${\mathbb R}^n$.
Once these notions are adopted, the Lorentz cone naturally appears as the cone of positive semidefinite
elements and the second symmetric function as determinant. For this reason, the Lorentz cones are
particularly well adapted to the study of concurrence. In this paper we never use the Jordan algebra
explicitly, so the reader is not required to be familiar with this concept.

Initially the concurrence is defined only for density matrices, but by homogeneity it can be extended to the whole
cone of positive semidefinite matrices \cite{Uhlmann05}. Once the determinant is defined on ${\mathbb R}^n$, we can generalize
the notion of concurrence in the following manner. Let $K \subset E$ be a convex cone defined in some real vector space $E$,
and let $\Gamma$ be the set of its generators, i.e.\ points which lie on extreme rays of $K$. Consider a linear map
$\Psi: E \to {\mathbb R}^n$ such that $\Psi[K] \subset L_n$. We call such maps $K$-to-$L_n$ positive.
Then we can define the concurrence
$C(\Psi;\cdot): K \to {\mathbb R}$ of $\Psi$ as the convex roof of the function $f(\gamma) = 2\sqrt{\det(\Psi(\gamma))}$.
Here $f$ is defined on $\Gamma$ and $\det$ is the determinant in ${\mathbb R}^n$ defined with respect to the Jordan
structure associated with the Lorentz cone $L_n$. We will show that this definition generalizes the notion of
concurrence as defined in \cite{Uhlmann05} in the following sense. For any positive map $\Phi$ there exists
$n \in {\mathbb N}_+$ and a map $\Psi$ sharing the input space $E$ with $\Phi$ and having output space ${\mathbb R}^n$
such that $C(\Phi;\cdot) = C(\Psi;\cdot)$. The role of the cone $K$ in the definition of the concurrence of $\Psi$
is played by the cone of positive semidefinite matrices. Similar definitions and results can be obtained for the $I$-fidelity.

If the input space $E$ of $\Psi$ is ${\mathbb R}^m$ and the cone $K$ is the Lorentz cone $L_m$, then we are able to
compute the concurrence $C(\Psi;\cdot)$ and the $I$-fidelity $F(\Psi;\cdot)$ explicitly. Since the Lorentz cone $L_4$ is isomorphic to the cone of
positive semidefinite complex hermitian $2 \times 2$ matrices, this allows us to compute the concurrence
and the $I$-fidelity of a positive map $\Phi$ whenever its input space is the space ${\cal H}(2)$ of  complex hermitian $2 \times 2$ matrices.
This, in turn, yields explicit formulae for the concurrence and the $I$-fidelity of bipartite density matrices of rank two. As an application,
we compute the concurrences and the $I$-fidelities of the density matrices of all graphs with 2 edges, as defined in \cite{Braunstein06}.

The paper is structured as follows. In the next section, we provide the necessary definitions.
We recall the definitions of concurrence as in \cite{Rungta01} and \cite{Uhlmann05}
and provide similar definitions for the $I$-fidelity. We introduce the Lorentz cones and define the necessary functions related to
their Jordan structure. Then we generalize the notions of concurrence and $I$-fidelity to relate them to this structure.
In section 3 we investigate the relation between bipartite matrices and completely positive maps and show that the concurrence
and $I$-fidelity of positive maps can be reduced to the concurrence and $I$-fidelity of some $K$-to-$L_n$ positive map.
Sections 4 and 5 contain the main theorems of the
paper, namely the formulae for the concurrence and the $I$-fidelity in the case when the input space of the $K$-to-$L_n$ positive map is
${\mathbb R}^m$ and $K$ is the Lorentz cone $L_m$. In section 6 we concretize these
results to the case of positive maps with input space ${\cal H}(2)$ and bipartite matrices of rank two.
In the next section we apply these results to the density matrices of all graphs with 2 edges. Finally, we summarize our results
and draw some conclusions in the last section.

\section{Definitions and preliminaries}

For some vector space $E$, let $\id_E$ be the identity operator on $E$. Denote by $I_n$ the $n \times n$ identity matrix, by $i$
the imaginary unit of the complex numbers, by $\interior W$ the interior of a set $W$ and by $\partial W$ its boundary.

For an $n \times n$ matrix $A$, denote by $\sigma_2(A)$
its second symmetric function $\sum_{1 \leq k < l \leq n} \lambda_k \lambda_l$, where $\lambda_1,\dots,\lambda_n$ are the eigenvalues
of $A$. The second symmetric function can be written as
\[ \sigma_2(A) = \frac{1}{2} \sum_{1 \leq k \not= l \leq n} \lambda_k \lambda_l =
\frac{1}{2} \left( \left( \sum_{k=1}^n \lambda_k \right)^2 - \sum_{k=1}^n \lambda_k^2 \right) =
\frac{1}{2} \left( (tr\,A)^2 - tr(A^2) \right).
\]

Let ${\cal H}(d)$ be the space of complex hermitian matrices of size $d \times d$ and let $H_+(d) \subset {\cal H}(d)$
be the cone of positive semidefinite matrices. The space ${\cal H}(d)$ has $d^2$ real dimensions.
We equip the second symmetric function on ${\cal H}(d)$ with an additional upper index $d$ to indicate the size
of its input matrices.
If $A \in {\cal H}(d)$, then $tr(A^2) = \langle A,A \rangle$ is the squared Frobenius norm of $A$.
Hence the second symmetric function on ${\cal H}(d)$ becomes
\begin{equation} \label{sigma2}
\sigma_2^d(A) = \frac{1}{2} \left( \langle I_d,A \rangle^2 - \sum_{k=1}^{d^2} \langle M_k,A \rangle^2 \right),
\end{equation}
where $M_1,\dots,M_{d^2}$ is any orthonormal basis of the space ${\cal H}(d)$.

A linear operator $\Phi: {\cal H}(d_1) \to {\cal H}(d_2)$ is said to be {\sl positive} if $\Phi[H_+(d_1)] \subset H_+(d_2)$.
It is said to be {\sl completely positive} if for any $n \in {\mathbb N}_+$, the operator $\id_{{\cal H}(n)} \otimes \Phi:
{\cal H}(n) \otimes {\cal H}(d_1) \to {\cal H}(n) \otimes {\cal H}(d_2)$ is positive.
The maximal rank achieved by matrices in the image of $\Phi$ is called the {\sl rank} of $\Phi$.

A completely positive operator can always be represented as a sum
\[ \Phi(X) = \sum_{k=1}^N A_k X A_k^*,
\]
where the {\sl Kraus operators} $A_1,\dots,A_N$ are complex $d_2 \times d_1$ matrices. The minimum number $N$ necessary for such a representation
of $\Phi$ is called the {\sl length} of $\Phi$ \cite{Uhlmann05}.

Fix two numbers $d_1,d_2 \in {\mathbb N}_+$. A $d_1 \otimes d_2$ {\sl bipartite matrix} $M$ is a matrix in ${\cal H}(d_1d_2)$
equipped with a block structure, namely partitioned into $d_1 \times d_1$ blocks $M_{kl}$ of size $d_2 \times d_2$ each,
where the indices $k,l$ run through $1,\dots,d_1$. Positive semidefinite $d_1 \otimes d_2$ bipartite matrices with trace 1
describe the mixed state of a composite quantum system consisting of a subsystem having $d_1$ states and a subsystem
having $d_2$ states. The {\sl partial trace} $tr_1$ of a $d_1 \otimes d_2$ bipartite matrix $M$ with respect to the first subsystem
is the $d_2 \times d_2$ matrix $\sum_{k=1}^{d_1} M_{kk}$. The partial trace $tr_2$ with respect to the second
subsystem is the $d_1 \times d_1$ matrix having $tr\,M_{kl}$ as $(k,l)$-entry.

The following definition of concurrence for positive operators is from \cite{Uhlmann05}.

{\definition \label{conc_ord} The {\sl concurrence} $C(\Phi;\cdot)$ of a positive operator $\Phi: {\cal H}(d_1) \to {\cal H}(d_2)$
is a real-valued function on the cone $H_+(d_1)$, defined as the largest convex function which coincides with
$2\sqrt{\sigma_2^{d_2}(\Phi(X))}$ on all rank 1 matrices $X \in H_+(d_1)$. }

\smallskip

We now introduce the following similar notion.

{\definition \label{fid_ord} The {\sl $I$-fidelity} $F(\Phi;\cdot)$ of a positive operator $\Phi: {\cal H}(d_1) \to {\cal H}(d_2)$
is a real-valued function on the cone $H_+(d_1)$, defined as the smallest concave function which coincides with
$2\sqrt{\sigma_2^{d_2}(\Phi(X))}$ on all rank 1 matrices $X \in H_+(d_1)$. }

\smallskip

By concretizing the positive operator $\Phi$ to be the partial trace, we can define these notions for
bipartite matrices. The following definition is essentially from \cite{Rungta01}.

{\definition \label{conc_bip} The {\sl concurrence} $C(M)$ of a positive semidefinite $d_1 \otimes d_2$ bipartite matrix $M$
is defined as $C(tr_1;M)$, i.e.\ $C(\cdot)$ is the largest convex function on $H_+(d_1d_2)$ which coincides with
$2\sqrt{\sigma_2^{d_2}(tr_1(X))}$ on all rank 1 matrices $X \in H_+(d_1d_2)$. }

\smallskip

Similarly we introduce the following notion.

{\definition \label{fid_bip} The {\sl $I$-fidelity} $F(M)$ of a positive semidefinite $d_1 \otimes d_2$ bipartite matrix $M$
is defined as $F(tr_1;M)$, i.e.\ $F(\cdot)$ is the largest convex function on $H_+(d_1d_2)$ which coincides with
$2\sqrt{\sigma_2^{d_2}(tr_1(X))}$ on all rank 1 matrices $X \in H_+(d_1d_2)$. }

\smallskip

Since for any positive semidefinite bipartite rank 1 matrix $\xi\xi^*$ we have $\sigma_2^{d_2}(tr_1(\xi\xi^*)) = \sigma_2^{d_1}(tr_2(\xi\xi^*))$,
we arrive at the same definitions if the partial trace is taken with respect to the second subsystem \cite{Rungta01}.
Here $\xi \in {\mathbb C}^{d_1d_2}$ denotes a vector.

\medskip

We now turn the the Lorentz cones and the associated Jordan structure. Let $e_0,e_1,\dots,e_{n-1}$ be the standard orthonormal basis vectors of ${\mathbb R}^n$.
For a vector $x \in {\mathbb R}^n$, let $x_0,x_1,\dots,x_{n-1}$ be the components of $x$ with respect to this basis. Let further $J_n$ be a diagonal matrix
whose first diagonal element equals 1 and whose all other diagonal elements equal $-1$.

The the {\sl Lorentz cone} $L_n \subset {\mathbb R}^n$ is defined as
\begin{equation} \label{deflor}
L_n = \left\{ x = (x_0,\dots,x_{n-1})^T \in {\mathbb R}^n \,|\, x_0 \geq \sqrt{\sum\limits_{k=1}^{n-1} x_k^2} \right\}.
\end{equation}
The Lorentz cone can be viewed as the cone of squares for a certain Jordan algebra. From this Jordan algebra the spaces ${\mathbb R}^n$
inherits the following structure \cite{KoecherBraun}.

{\definition \label{Jordandef} For any vector $x = (x_0,\dots,x_{n-1})^T \in {\mathbb R}^n$, let $\lambda_{\pm} = x_0 \pm \sqrt{\sum_{k=1}^{n-1} x_k^2}$ be the {\sl eigenvalues} of $x$.
Define further $\lambda_+ + \lambda_- = 2x_0$ to be the {\sl trace} $tr\,x$ and $\lambda_+\lambda_- = x_0^2 - \sum_{k=1}^{n-1} x_k^2$ to be the
{\sl determinant} $\det x$ of $x$. }

\smallskip

With these definitions, a vector $x$ is contained in $L_n$ if and only if its eigenvalues are nonnegative, and it is contained in the interior of $L_n$
if its eigenvalues are positive. Hence the Lorentz cone can be seen as an analogue of a $2 \times 2$ positive semidefinite matrix cone.
The determinant of a vector $x$ can be simply written as
\[ \det x = x^TJ_nx.
\]

We now define the linear map ${\cal I}: {\mathbb R}^4 \to {\cal H}(2)$ by
\[ {\cal I}(x) = \left( \begin{array}{cc} x_0+x_1 & x_2+ix_3 \\ x_2-ix_3 & x_0-x_1 \end{array} \right).
\]
As can easily be verified, the eigenvalues of the matrix ${\cal I}(x)$ equal the eigenvalues of $x$ for any $x \in {\mathbb R}^4$,
and ${\cal I}[L_4] = H_+(2)$. Thus $L_4$ can actually be identified with $H_+(2)$ by virtue of the isomorphism ${\cal I}$.

{\remark In the same manner, the cone $L_3$ is isomorphic to the cone of positive semidefinite $2 \times 2$ real symmetric matrices
and the cone $L_6$ to the cone of positive semidefinite $2 \times 2$ quaternionic hermitian matrices. Therefore the formulae obtained in this paper
equally apply for positive maps having as input space the space of real symmetric or quaternionic hermitian $2 \times 2$ matrices,
given the definitions of concurrence and $I$-fidelity are adapted accordingly. }

\medskip

We now extend the notions of concurrence and $I$-fidelity to ${\mathbb R}^n$-valued positive maps.
Let $E$ be a real vector space and $K \subset E$ be a regular (i.e.\ closed, containing no lines) convex cone. The following notion is standard in convex analysis \cite{Rockafellar}.

{\definition A point $y \in K$ is said to be {\sl extremal} if $y_1,y_2 \in K$, $y = y_1 + y_2$ implies $y_1 = \alpha_1 y$, $y_2 = \alpha_2 y$
for some nonnegative scalars $\alpha_1,\alpha_2$. }

\smallskip

The extremal points of the cone $H_+(d)$ of positive semidefinite matrices are precisely the matrices of the form $\xi\xi^*$, where $\xi \in {\mathbb C}^d$.
The extremal points of the cone $L_n$ are precisely the points lying on the boundary $\partial L_n$.

{\definition Assume above notations and let $m,n \in {\mathbb N}_+$ be some integers. We call a linear map $\Psi: E \to {\mathbb R}^n$ {\sl $K$-to-$L_n$ positive}
if $\Psi[K] \subset L_n$. We call a linear map $\Psi: {\mathbb R}^m \to {\mathbb R}^n$ {\sl Lorentz-positive}
if $\Psi[L_m] \subset L_n$.}

{\definition \label{conc_lor} Let $\Gamma$ be the set of extremal points of a regular convex cone $K \subset E$, where $E$ is a real vector space. 
Let $\Psi: E \to {\mathbb R}^n$ be a $K$-to-$L_n$ positive map,
where $n \in {\mathbb N}_+$ is some integer. The {\sl concurrence} of $\Psi$ is the largest convex function
$C(\Psi;\cdot)$ on the cone $K$ which coincides with the function $f(\gamma) = 2\sqrt{\det(\Psi(\gamma))}$ on all extremal points $\gamma \in \Gamma$. }

{\definition \label{fid_lor} Assume the notations of the previous definition. The {\sl $I$-fidelity} of $\Psi$ is the smallest concave function
$F(\Psi;\cdot)$ on the cone $K$ which coincides with the function $f(\gamma) = 2\sqrt{\det(\Psi(\gamma))}$ on all extremal points $\gamma \in \Gamma$. }

\smallskip

We thus define concurrence and $I$-fidelity as a convex and concave roof, respectively. In comparison to
Definitions \ref{conc_ord}, \ref{fid_ord}, we do not restrict the input space of the operator $\Psi$ to be a
space of self-adjoint operators, but in contrast we restrict its output space to be ${\mathbb R}^n$ equipped
with a corresponding Jordan structure. This allows us to replace the second symmetric function $\sigma_2$ by
the determinant. As we will see further, this does in fact not at all restrict the generality.

\section{Relations between the different definitions}

In this section we investigate the relation between the different definitions of concurrence and
$I$-fidelity.

\medskip

First we explore the connection between the concurrence or $I$-fidelity of completely positive maps and that of bipartite matrices.

Let $\Phi: {\cal H}(d_1) \to {\cal H}(d_2)$ be a completely positive map, with Kraus representation
\[ \Phi(X) = \sum_{k=1}^{d_3} A_kXA_k^*.
\]
With this representation we associate the third order tensor ${\bf A}_{\alpha\beta\gamma}$, of dimension $d_1 \times d_2
\times d_3$ and with elements ${\bf A}_{\alpha\beta\gamma} = (A_{\gamma})_{\beta\alpha}$, $\alpha =
1,\dots,d_1$, $\beta = 1,\dots,d_2$, $\gamma = 1,\dots,d_3$.
Further we associate to this representation the $d_2d_3 \times d_1$ matrix
\[ A = \left( \begin{array}{c} A_1 \\ A_2 \\ \vdots \\ A_{d_3} \end{array} \right).
\]
Then $\Phi(X)$ equals the partial trace with respect to the first subsystem of the $d_3 \otimes d_2$ bipartite matrix $AXA^*$,
$\Phi(X) = tr_1(AXA^*)$. Therefore by Definitions \ref{conc_ord}, \ref{conc_bip} we have
\[ C(\Phi;X) = C(tr_1;AXA^*) = C(AXA^*).
\]
Now note that the concurrence of a bipartite matrix is independent of which subsystem is taken to define the partial trace, i.e.\
$C(tr_1;AXA^*) = C(tr_2;AXA^*)$. If we define another positive map $\Phi': {\cal H}(d_1) \to {\cal H}(d_3)$ by $\Phi'(X) = tr_2(AXA^*)$, then it follows that
\[ C(\Phi;X) = C(\Phi';X)
\]
for all $X \in H_+(d_1)$. In a similar manner we obtain the equality
\[ F(\Phi;X) = F(\Phi';X) \qquad \forall\ X \in H_+(d_1)
\]
for the $I$-fidelity.

The map $\Phi'$ has a Kraus representation given by
\begin{equation} \label{Kraus'}
\Phi'(X) = \sum_{k=1}^{d_2} A'_kX{A'_k}^*,
\end{equation}
where the element $(\gamma,\alpha)$ of the $d_3 \times d_1$ matrix $A'_{\beta}$ is given by $(A_{\gamma})_{\beta\alpha} = {\bf A}_{\alpha\beta\gamma}$.
Hence the third order tensor associated to the Kraus representation (\ref{Kraus'}) is given by ${\bf A}'_{\alpha\gamma\beta} = {\bf A}_{\alpha\beta\gamma}$
and has dimension $d_1 \times d_3 \times d_2$. The map $\Phi'$ can therefore be obtained from $\Phi$ by exchanging the last two indices
in the corresponding third order tensor ${\bf A}$. From the point of view of concurrence, the rank and the length of a completely positive map
are hence interchangeable. We obtain the following result.

{\lemma \label{pos_bip}
Let $\Phi$ be a completely positive map with input space ${\cal H}(d_1)$, rank $d_2$, and length $d_3$. Then there exists a completely positive map $\Phi'$
with input space ${\cal H}(d_1)$, rank $d_3$, and length $d_2$ such that
\[ C(\Phi;X) = C(\Phi';X),\quad F(\Phi;X) = F(\Phi';X) \qquad \forall\ X \in H_+(d_1).
\]
Moreover, for any matrix $X \in H_+(d_1)$ of rank $k \leq d_1$ there exists a $d_2 \otimes d_3$ bipartite matrix $M_X$ of rank $k$ such that
\[ C(\Phi;X) = C(M_X),\quad F(\Phi;X) = F(M_X). \qedhere
\] }

The construction of $\Phi'$ and $M_X$ is quite obvious from the above. This lemma may be a reason why the concurrence of completely positive maps of rank and length two \cite{Uhlmann05}
eventually boils down to an analogue of Wootters formula for the concurrence of $2 \otimes 2$ bipartite matrices \cite{Wootters98}.

\medskip

We now explore the relation between Definitions \ref{conc_ord}, \ref{fid_ord} and \ref{conc_lor}, \ref{fid_lor}, respectively.

Let $\Phi: {\cal H}(d_1) \to {\cal H}(d_2)$ be an arbitrary positive map. Let further $\{M_1,\dots,M_{d_2^2}\}$ be an orthonormal basis of the space ${\cal H}(d_2)$.
We then have by (\ref{sigma2})
\[ \sigma_2^{d_2}(\Phi(X)) = \frac{1}{2} \left((tr\,\Phi(X))^2 - \sum_{k=1}^{d_2^2} \langle \Phi(X), M_k \rangle^2 \right).
\]
Let us define a linear map $\Phi_L: {\cal H}(d_1) \to {\mathbb R}^{d_2^2+1}$ by
\begin{equation} \label{PhiLdef}
\Phi_L(X) = \frac{\sqrt{2}}{2} \left( \begin{array}{c} tr\,\Phi(X) \\ \langle \Phi(X), M_1 \rangle \\ \vdots \\ \langle \Phi(X), M_{d_2^2} \rangle \end{array} \right).
\end{equation}
Note that since $\Phi$ is a positive map, we have $\Phi(X) \succeq 0$ and hence $tr\,\Phi(X) \geq ||\Phi(X)||_2$ for all $X \succeq 0$.
Therefore the linear map $\Phi_L$ takes the positive semidefinite matrix cone $H_+(d_1)$ to the Lorentz cone $L_{d_2^2+1}$
and is hence a $H_+(d_1)$-to-$L_{d_2^2+1}$ positive map.
With Definition \ref{Jordandef} we then have
\begin{equation} \label{equal_formula}
\sigma_2^{d_2}(\Phi(X)) = \Phi_L(X)^T J_{d_2^2+1} \Phi_L(X) = \det\Phi_L(X).
\end{equation}
The concurrences of the maps $\Phi,\Phi_L$ are defined by Definitions \ref{conc_ord}, \ref{conc_lor}, respectively. In view of the above equation they are the convex roofs of the same function.
It follows that
\[ C(\Phi;X) = C(\Phi_L;X) \qquad \forall\ X \in H_+(d_1).
\]
In a similar manner we obtain
\[ F(\Phi;X) = F(\Phi_L;X) \qquad \forall\ X \in H_+(d_1)
\]
for the $I$-fidelities, defined by Definitions \ref{fid_ord}, \ref{fid_lor}, respectively. We obtain the following result.

{\lemma \label{equal_functions}
Let $\Phi: {\cal H}(d_1) \to {\cal H}(d_2)$ be a positive map. Then there exists a $H_+(d_1)$-to-$L_{d_2^2+1}$ positive map $\Phi_L$
such that
\[ C(\Phi;X) = C(\Phi_L;X),\quad F(\Phi;X) = F(\Phi_L;X) \qquad \forall\ X \in H_+(d_1).
\]
The map $\Phi_L$ can be constructed as in (\ref{PhiLdef}). \qed }

The lemma tells us that Definitions \ref{conc_lor}, \ref{fid_lor} are actually generalizations of Definitions \ref{conc_ord}, \ref{fid_ord}.

\section{Concurrence of Lorentz-positive maps}

There exists a simple case when we are able to compute the concurrence of a $K$-to-$L_n$ positive map explicitly, namely when the cone $K$ in the input space is also a Lorentz cone.
In this section we will derive a formula for the concurrence of Lorentz-positive maps.

\medskip

Let us first investigate the properties of certain real symmetric matrix pencils related to Lorentz-positive maps.

{\lemma Let $n \geq 3$ be an integer and suppose that $P = P^T$, $J = J^T$ are real symmetric $n \times n$ matrices, $J$ is regular with signature $(+-\cdots-)$ and
there exists a number $\hat\lambda \in {\mathbb R}$ such that $P - \hat\lambda J \succ 0$. 
Let $\lambda_1,\dots,\lambda_n$ be the generalized eigenvalues of the matrix pencil $P - \lambda J$, with their real parts in decreasing order.
Then the following assertions hold.
\begin{tabbing}
(iii)\ \= \kill
(i) \> all eigenvalues $\lambda_1,\dots,\lambda_n$ are real, \\
(ii) \> $\lambda_1 \not= \lambda_2$ and the matrix $P - \lambda J$ is positive definite if and only if $\lambda_2 < \lambda < \lambda_1$, \\
(iii) \> if $\lambda < \lambda_1$, then $x^T(P - \lambda J)x \leq 0$ implies $x^TJx \leq 0$ for any $x \in {\mathbb R}^n$.
\end{tabbing} }

\begin{proof}
Since $J$ is invertible, the generalized eigenvalues of the matrix pencil $P - \lambda J$ are the eigenvalues of the matrix $J^{-1}P$
and there are indeed exactly $n$ of them.

Since the signatures of $J$ and $-J$ are different, there must be at least one real number $\lambda$
such that $P - \lambda J$ is singular, i.e.\ at least one generalized eigenvalue of the matrix pencil must be real.
Denote the maximal real eigenvalue of $J^{-1}P$ by $\lambda_{\max}$ and the minimal real eigenvalue by $\lambda_{\min}$.
Then for $\lambda > \lambda_{\max}$ the matrix $P - \lambda J$ has the same signature as $-J$, and
for $\lambda < \lambda_{\min}$ it has the same signature as $J$.
It follows that the signatures of $P - \hat\lambda J$ and $P - \lambda J$ differ by exactly one sign for any $\lambda > \lambda_{\max}$
and by exactly $n-1$ signs for any $\lambda < \lambda_{\min}$. Therefore the interval $(\hat\lambda,\lambda_{\max}]$ contains
at least 1 eigenvalue and the interval $[\lambda_{\min},\hat\lambda)$ at least $n-1$ eigenvalues.

Thus all eigenvalues must be real, $\lambda_{\max} = \lambda_1$, $\lambda_{\min} = \lambda_n$ and the interval $(\hat\lambda,\lambda_1)$
does not contain any eigenvalue. This proves (i), and (ii) follows by convexity of the cone of positive semidefinite matrices.

Let us prove (iii). Assume the contrary, i.e.\ that there exists $\lambda < \lambda_1$ and $x \in {\mathbb R}^n$ such that $x^TJx > 0$ and $x^T(P - \lambda J)x \leq 0$.
Then we have $x^T(P - \lambda_1 J)x = x^T(P - \lambda J)x + (\lambda - \lambda_1) x^TJx < 0$.
But $P - \lambda_1 J \succeq 0$ by (ii), which leads to a contradiction and completes the proof.
\end{proof}

We actually need the following version of this lemma with relaxed assumptions.

{\lemma \label{eigvpencil} Let $n \geq 3$ be an integer and suppose that $P = P^T$, $J = J^T$ are real symmetric $n \times n$ matrices, $J$ is regular with signature $(+-\cdots-)$ and
there exists a number $\hat\lambda \in {\mathbb R}$ such that $P - \hat\lambda J \succeq 0$.
Let $\lambda_1,\dots,\lambda_n$ be the generalized eigenvalues of the matrix pencil $P - \lambda J$, with their real parts in decreasing order.
Then the following assertions hold.
\begin{tabbing}
(iii)\ \= \kill
(i) \> all eigenvalues $\lambda_1,\dots,\lambda_n$ are real, \\
(ii) \> the matrix $P - \lambda J$ is positive semidefinite if and only if $\lambda_2 \leq \lambda \leq \lambda_1$, \\
(iii) \> if $\lambda < \lambda_1$, then $x^T(P - \lambda J)x \leq 0$ implies $x^TJx \leq 0$ for any $x \in {\mathbb R}^n$.
\end{tabbing} }

\begin{proof}
Assertions (i) and (ii) can be obtained by continuity arguments from the previous lemma
when replacing $P$ with $P + \varepsilon A$ for any $\varepsilon > 0$, $A \succ 0$, and taking the limit $\varepsilon \to 0$.
Assertion (iii) is proven the same way as in the previous lemma.
\end{proof}


\smallskip

The following assertion is a well-known consequence of the ${\cal S}$-lemma \cite{Dines43},\cite{Yakubovich71}.

{\lemma \label{Slemma} Let $\Upsilon: {\mathbb R}^m \to {\mathbb R}^n$ be a Lorentz-positive map, represented by an $n \times m$ matrix.
Then there exists $\hat\lambda \geq 0$ such that such that $\Upsilon^TJ_n\Upsilon \succeq \hat\lambda J_m$. \qed }

A proof is for instance in \cite[Lemma 1]{Hildebrand0503194}.

\medskip

This lemma relates the Lorentz-positive map $\Upsilon$ to the real symmetric matrix pencil $\Upsilon^TJ_n\Upsilon - \lambda J_m$.
By Lemma \ref{eigvpencil} this matrix pencil has $m$ real eigenvalues. Let $\lambda_1,\dots,\lambda_m$ denote these eigenvalues in decreasing order.
By assertion (ii) of Lemma \ref{eigvpencil} $\lambda_2$ is the smallest $\lambda \in {\mathbb R}$ such that $\Upsilon^TJ_n\Upsilon \succeq \lambda J_m$.
Hence by a convex separation argument there exists a nonzero $\hat x \in {\mathbb R}^m$ such that $\hat x^T(\Upsilon^TJ_n\Upsilon - \lambda_2 J_m)\hat x = 0$
and $\hat x^TJ_m\hat x \leq 0$. As a consequence, $\hat x$ will be linearly independent from any vector in the interior of $L_m$, since $y^TJ_my > 0$ for any $y \in \interior L_m$.

We shall now compute the concurrence of $\Upsilon$, as given by Definition \ref{conc_lor}.
Denote the positive semidefinite matrix $\Upsilon^TJ_n\Upsilon - \lambda_2 J_m$ by $Q$.
For any vector $x \in \partial L_m$ we have $x^TJ_mx = 0$ and hence
\[ C(\Upsilon;x) = 2\sqrt{\det \Upsilon(x)} = 2\sqrt{x^T\Upsilon^TJ_n\Upsilon x} = 2\sqrt{x^T (\Upsilon^TJ_n\Upsilon - \lambda_2 J_m) x} = 2\sqrt{x^T Q x}.
\]
The concurrence $C(\Upsilon;\cdot): L_m \to {\mathbb R}$ is then the largest convex function on $L_m$ that coincides with the
expression above on all vectors $x \in \partial L_m$, i.e.\ its convex roof (see \cite{Uhlmann00}).
Let us now show that the function $2\sqrt{x^TQx}$ {\it is} actually this convex roof.

For any $x \in {\mathbb R}^m$ such that $x^TQx > 0$ we have
\[ \frac{\partial^2 \sqrt{x^TQx}}{\partial x^2} = \frac{Q(x^TQx) - Qxx^TQ}{(x^TQx)^{3/2}}.
\]
If we evaluate this Hessian on the vector $y$, we get $\frac{(y^TQy)(x^TQx) - (y^TQx)^2}{(x^TQx)^{3/2}}$.
Since $Q \succeq 0$, we have $(y^TQy)(x^TQx) \geq (y^TQx)^2$ by the Cauchy-Schwarz inequality for the degenerate scalar product defined by $Q$
and the Hessian is positive semidefinite.
Thus the function $2\sqrt{x^TQx}$ is convex in a convex neighbourhood $U_x$ of any vector $x$ such that $x^TQx > 0$.
However, if $x^TQx = 0$ for some vector $x$, then the constant zero function supports $2\sqrt{x^TQx}$ at $x$. It follows that the function
$2\sqrt{x^TQx}$ is convex on the whole space ${\mathbb R}^m$.

Let us now prove that $2\sqrt{x^TQx}$ is a roof (for a definition see \cite{Uhlmann00}). For any vector
$\tilde x \in \interior L_m$, define $L_{\tilde x}$ as the 2-dimensional subspace $L_{\tilde x} \subset
{\mathbb R}^m$ spanned by $\tilde x,\hat x$. Since $Q\hat x = 0$ as a consequence of the relation $\hat
x^TQ\hat x = 0$ and the positivity of $Q$, we have for any $\alpha,\beta \in {\mathbb R}$ and $x = \alpha
\tilde x + \beta \hat x$ that $2\sqrt{x^TQx} = 2\sqrt{\alpha^2 \tilde x^T Q \tilde x} = 2|\alpha|\sqrt{\tilde x^T
Q \tilde x}$. Hence the restriction of the function $2\sqrt{x^TQx}$ on the set $S_{\tilde x} = \{ x = \alpha
\tilde x + \beta \hat x \,|\, \alpha \geq 0 \}$ is linear.

We shall now prove that $S_{\tilde x} \cap L_m = L_{\tilde x} \cap L_m$. Assume the contrary, i.e.\ that there exists
$\alpha < 0$, $\beta \in {\mathbb R}$ such that $\alpha \tilde x + \beta \hat x \in L_m$. Then $\beta \not= 0$,
because $\alpha \tilde x \in -\interior L_m$. Moreover, $-\alpha \tilde x \in \interior L_m$ and by convexity of $L_m$
we have $\frac{1}{2}(-\alpha \tilde x + (\alpha \tilde x + \beta \hat x)) = \frac{\beta}{2} \hat x \in \interior L_m$.
But $\hat x \not= \pm\interior L_m$, which leads to a contradiction.

Therefore any vector $\tilde x \in \interior L_m$ can be
represented as a convex combination of two points in $\partial L_m \cap S_{\tilde x}$. Hence the function
$2\sqrt{x^TQx}$ is indeed a roof.

Note that $x^T\Upsilon^TJ_n\Upsilon x = \det \Upsilon(x)$ and $x^TJ_mx = \det x$. We have proven the following result.

{\theorem \label{concl2} Let $\Upsilon: {\mathbb R}^m \to {\mathbb R}^n$ be a Lorentz-positive map.
Then the pencil $\det \Upsilon(x) - \lambda \det x$ of quadratic forms on ${\mathbb R}^m$
has $m$ real generalized eigenvalues. Let $\lambda_2$ be the second largest of them. Then the concurrence of $\Upsilon$
is given by the expression
\[ C(\Upsilon;x) = 2\sqrt{ \det \Upsilon(x) - \lambda_2 \det x }
\]
for all $x \in L_m$.

The function $C(\Upsilon;\cdot)$ is linear along any affine subspace which is parallel to the eigenspace $V$ corresponding to the eigenvalue $\lambda_2$ of the pencil
$\det \Upsilon(x) - \lambda \det x$. For any $x \in \interior L_m$
there exists a vector in $V$ which is linearly independent of $x$. This allows to obtain an optimal
decomposition of $x$ as convex combination $x = \mu y + (1-\mu) z$ of two points $y,z \in \partial L_m$
such that $C(\Upsilon;x) = \mu C(\Upsilon;y) + (1-\mu) C(\Upsilon;z)$.
\qed }

\section{$I$-fidelity of Lorentz-positive maps}

In this subsection we prove a result analogous to Theorem \ref{concl2} for the $I$-fidelity of Lorentz-positive maps.
Assume the notations of the previous section.

{\theorem \label{fidlm} Let $\Upsilon: {\mathbb R}^m \to {\mathbb R}^n$ be a Lorentz-positive map.
Then the pencil $\det \Upsilon(x) - \lambda \det x$ of quadratic forms on ${\mathbb R}^m$
has $m$ real generalized eigenvalues. Let $\lambda_{\min}$ be the smallest of them. Then the $I$-fidelity of $\Upsilon$
is given by the expression
\[ F(\Upsilon;x) = 2\sqrt{ \det \Upsilon(x) - \lambda_{\min} \det x }
\]
for all $x \in L_m$.

The function $F(\Upsilon;\cdot)$ is linear along any affine subspace which is parallel to the eigenspace $V$ corresponding to the eigenvalue $\lambda_{\min}$ of the pencil
$\det \Upsilon(x) - \lambda \det x$. For any $x \in \interior L_m$
there exists a vector in $V$ which is linearly independent of $x$. This allows to obtain an optimal
decomposition of $x$ as convex combination $x = \mu y + (1-\mu) z$ of two points $y,z \in \partial L_m$
such that $F(\Upsilon;x) = \mu F(\Upsilon;y) + (1-\mu) F(\Upsilon;z)$.}

\begin{proof}
Denote the generalized eigenvalues of the matrix pencil $\Upsilon^TJ_n\Upsilon - \lambda J_m$ by $\lambda_1,\dots,\lambda_m$, in decreasing order.
By Lemma \ref{Slemma} Lemma \ref{eigvpencil} is applicable and all generalized eigenvalues of the pencil are real.

Define the matrices $Q_m = \Upsilon^TJ_n\Upsilon - \lambda_m J_m$ and $Q_1 = \Upsilon^TJ_n\Upsilon - \lambda_1 J_m$.
The $I$-fidelity of $\Upsilon$ at a vector $x \in \partial L_m$ is given by
\[ F(\Upsilon;x) = 2\sqrt{\det \Upsilon(x)} = 2\sqrt{x^T\Upsilon^TJ_n\Upsilon x} = 2\sqrt{x^T (\Upsilon^TJ_n\Upsilon - \lambda_m J_m) x} = 2\sqrt{x^T Q_m x},
\]
because $x^TJ_mx = 0$ for any $x \in \partial L_m$.
The $I$-fidelity $F(\Upsilon;\cdot): L_m \to {\mathbb R}$ is then the smallest concave function on $L_m$ that coincides with the
expression above on all vectors $x \in \partial L_m$, i.e.\ its concave roof (see \cite{Uhlmann00}).
We have to show that the function $2\sqrt{x^TQ_mx}$ {\it is} actually this concave roof.

We have $Q_m = Q_1 + (\lambda_1 - \lambda_m)J_m$. Hence for any vector $x \in L_m$ we have $x^TQ_mx = x^TQ_1x + (\lambda_1 - \lambda_m)x^TJ_mx \geq 0$,
because $Q_1 \succeq 0$ by assertion (ii) of Lemma \ref{eigvpencil}, $\lambda_1 \geq \lambda_m$ and $x^TJ_mx \geq 0$. Hence $2\sqrt{x^TQ_mx}$ is well-defined on $L_m$.

If $\lambda_1 = \lambda_m$, then all generalized eigenvalues of the pencil are equal and the polynomial $p(\lambda) = \det(\Upsilon^TJ_n\Upsilon - \lambda J_m)$
has an $m$-fold root at $\lambda = \lambda_m$. Hence the matrix $Q_m$ must be zero, in which case $2\sqrt{x^TQ_mx} \equiv 0$ is a concave roof.

Let us henceforth assume that $\lambda_1 > \lambda_m$. Let $\hat x$ be an eigenvector to the eigenvalue $\lambda_m$ of the pencil $\Upsilon^TJ_n\Upsilon - \lambda J_m$.
Then $Q_m\hat x = 0$, $\hat x^TQ_m\hat x = 0$ and $\hat x^TJ_m\hat x \leq 0$ by assertion (iii) of Lemma \ref{eigvpencil}. It follows that $x,\hat x$ are linearly
independent for any $x \in \interior L_m$.

Let us show that $2\sqrt{x^TQ_mx}$ is concave on $L_m$. For any $\varepsilon > 0$ the signature of the matrix $Q_m + \varepsilon J_m$ equals that of $J_m$,
because $Q_m + \varepsilon J_m$ is regular for any $\varepsilon > 0$. Consequently, $Q_m$ has at most one positive eigenvalue and
we can represent $Q_m$ as $S^TJ_mS$, where $S$ is some singular real $m \times m$ matrix.
Let now $x \in \interior L_m$. Then we have $x^TJ_mx > 0$ and $x^TQ_mx = x^TQ_1x + (\lambda_1 - \lambda_m)x^TJ_mx > 0$.
As in the previous section, it follows that
\[ \frac{\partial^2 \sqrt{x^TQ_mx}}{\partial x^2} = \frac{S^TJ_mS \cdot x^TS^TJ_mSx - S^TJ_mSxx^TS^TJ_mS}{(x^TQ_mx)^{3/2}}.
\]
Let us evaluate this Hessian on the vector $y$. Denote $Sx$ by $\bar x$ and $J_mSy$ by $\bar y$. Then we get
\[ y^T\frac{\partial^2 \sqrt{x^TQ_mx}}{\partial x^2}y = \frac{\bar y^TJ_m\bar y \cdot \bar x^TJ_m\bar x - \bar y^T\bar x\bar x^T\bar y}{(x^TQ_mx)^{3/2}}
= -\frac{\bar y^T(\bar x\bar x^T - \bar x^TJ_m\bar x \cdot J_m)\bar y}{(x^TQ_mx)^{3/2}}.
\]
Note that $\bar x^TJ_m\bar x = x^TQ_mx > 0$ and hence $\bar x \in \pm\interior L_m$. It follows that $J_m \bar x
\in \pm\interior L_m$. But $J_m\bar x$ is an eigenvector of the quadratic form $\bar x\bar x^T - \bar x^TJ_m\bar
x \cdot J_m$ with eigenvalue zero.

Assume that this form has an eigenvector $z$ with negative eigenvalue. Then $z$ is orthogonal to $J_m\bar x$ and hence
$z^T J_m z < 0$. This is because a vector in $\interior L_m$ cannot be orthogonal to any other nonzero vector in $L_m$.
But then $z^T(\bar x\bar x^T - \bar x^TJ_m\bar x \cdot J_m)z = (\bar x^Tz)^2 + (\bar x^TJ_m\bar
x)(-z^T J_m z) \geq 0$, which leads to a contradiction.

Thus $\bar x\bar x^T - \bar x^TJ_m\bar x \cdot J_m \succeq 0$ and
$\frac{\partial^2 \sqrt{x^TQ_mx}}{\partial x^2} \preceq 0$. It follows that the function
$2\sqrt{x^TQ_mx}$ is concave on $\interior L_m$ and hence by continuity on the whole cone $L_m$.

By the same arguments as for the proof of Theorem \ref{concl2}, any vector $\tilde x \in \interior L_m$ can be represented as a
convex combination of two points in $\partial L_m \cap L_{\tilde x} = \partial L_m \cap S_{\tilde x}$,
where $L_{\tilde x},S_{\tilde x}$ are defined as in the previous section and the
restriction of the function $2\sqrt{x^TQ_mx}$ on the set $S_{\tilde x}$ is linear. Thus $2\sqrt{x^TQ_mx}$ is
indeed a roof.
\end{proof}

\section{Concurrence and $I$-fidelity of positive maps}

In this section we apply the results of the previous sections to compute the concurrence and the $I$-fidelity
of positive maps with input space ${\cal H}(2)$ and bipartite density matrices of rank 2.

As we have seen in Section 2, the cone $H_+(2)$ of positive semidefinite complex hermitian $2 \times 2$ matrices
is isomorphic to $L_4$, with the matrix determinant being equal to the determinant induced on ${\mathbb R}^4$
by the Jordan structure of $L_4$. By application of Lemma \ref{equal_functions} and relation (\ref{equal_formula})
Theorems \ref{concl2} and \ref{fidlm} then yield the following results.

{\theorem \label{pos_result} Let $\Phi: {\cal H}(2) \to {\cal H}(d_2)$ be a positive operator.
Then the pencil $\sigma_2^{d_2} (\Phi(X)) - \lambda \det X$ of quadratic forms on ${\cal H}(2)$
has 4 real generalized eigenvalues. Denote these eigenvalues by $\lambda_1,\lambda_2,\lambda_3,\lambda_4$, in decreasing order.
Then the concurrence and the $I$-fidelity of $\Phi$ are given by the expressions
\[ C(\Phi;X) = 2\sqrt{ \sigma_2^{d_2} (\Phi(X)) - \lambda_2 \det X }, \qquad
F(\Phi;X) = 2\sqrt{ \sigma_2^{d_2} (\Phi(X)) - \lambda_4 \det X }
\]
for all $X \in H_+(2)$.

The function $C(\Phi;\cdot)$ (respectively, $F(\Phi;\cdot)$) is linear along any affine subspace which is parallel to
the eigenspace $V_2$ (respectively, $V_4$) corresponding to the eigenvalue $\lambda_2$ (respectively, $\lambda_4$) of the pencil
$\sigma_2^{d_2} (\Phi(X)) - \lambda \det X$.
For any $X \in \interior H_+(2)$
there exists a matrix in $V_2$ (respectively, $V_4$) which is linearly independent of $X$. This allows to obtain an
optimal decomposition of $X$ as convex combination $X = \mu Y + (1-\mu) Z$ of two rank 1 matrices $Y,Z \in H_+(2)$
such that $C(\Phi;X) = \mu C(\Phi;Y) + (1-\mu) C(\Phi;Z)$ (respectively, $F(\Phi;X) = \mu F(\Phi;Y) + (1-\mu) F(\Phi;Z)$).
\qed }

\smallskip

Note that the formulae provided by Theorems \ref{concl2} and \ref{fidlm} are coordinate independent.
Moreover, the formulae do not change if the quadratic form $\det x$ is scaled by multiplication with a positive number.
We are therefore not
bound to the standard Lorentz cone $L_n$ as defined in (\ref{deflor}). All we need are two quadratic forms
on the input space, one having signature $(+-\cdots-)$ and defining a convex cone by its zero set, and the other
describing the value of the concurrence or $I$-fidelity on the extremal elements of this cone.
In fact, the proofs of Theorems \ref{concl2} and \ref{fidlm} used the assumption that $\Upsilon$ is Lorentz-positive only
to ensure the assumptions of Lemma \ref{eigvpencil}. We can hence relax the Lorentz-positivity of $\Upsilon$
and even abstract ourselves from the notion of a positive map. We have the following general result, whose proof
goes along the lines of proof of Theorems \ref{concl2} and \ref{fidlm}, with obvious modifications.

{\theorem \label{main} Let $E$ be a real vector space of dimension $n$ and let $J$ be a regular quadratic form on $E$ with signature $(+-\cdots-)$.
The set $\{ x \in E \,|\, J(x) \geq 0 \}$ forms two convex cones, which are linearly isomorphic to the Lorentz cone $L_n$.
Let $K \subset E$ be one of these two cones. Let $P$ be another quadratic form on $E$ satisfying the condition
\begin{equation} \label{main_cond}
\exists\ \lambda \in {\mathbb R}: \qquad P \succeq \lambda J.
\end{equation}
Then the pencil $P - \lambda J$ of quadratic forms has $n$ real generalized eigenvalues. Let $\lambda_1,\dots,\lambda_n$
denote these eigenvalues in decreasing order. Let the function $p: \delta K \to {\mathbb R}$ be defined by $p(x) = 2\sqrt{P(x)}$.

Then the largest convex function $p_{conv}$ and the smallest concave function $p_{conc}$ on $K$ which coincide with $p$ on $\partial K$ are given by
\[ p_{conv}(x) = 2\sqrt{P(x) - \lambda_2 J(x)},\qquad p_{conc}(x) = 2\sqrt{P(x) - \lambda_n J(x)}.
\]
These two functions are hence the convex and the concave roof of $p$, respectively.

Let $V_2,V_n \subset E$ be the eigenspaces corresponding
to the eigenvalues $\lambda_2,\lambda_n$, respectively, of the pencil $P - \lambda J$. Then $p_{conv}$ is linear on any affine subspace
parallel to $V_2$ and $p_{conc}$ is linear on any affine subspace
parallel to $V_n$. For any $x \in \interior K$ there exist elements in $V_2$ and $V_n$ which are linearly independent of $x$.
This allows to obtain optimal decompositions of $x$ as convex combinations $x = \mu y_1 + (1-\mu) y_2 =
\eta z_1 + (1-\eta) z_2$ of points $y_1,y_2,z_1,z_2 \in \partial K$ such that
$p_{conv}(x) = \mu p(y_1) + (1-\mu) p(y_2)$, $p_{conc}(x) = \eta p(z_1) + (1-\eta) p(z_2)$.
\qed }

\smallskip

Let $d_1 \geq 2$ and let $P \subset {\mathbb C}^{d_1}$ be a linear complex subspace of dimension 2. Denote by $U_P$ the subspace of all matrices in ${\cal H}(d_1)$
whose range is contained in $P$. Then $U_P$ is isomorphic to ${\cal H}(2)$ and the intersection
$K_P = U_P \cap H_+(d_1)$ is isomorphic to $H_+(2)$ and hence to $L_4$. Moreover, if an element of $K_P$
is represented as convex combination of extremal elements of $H_+(d_1)$, then all these extremal elements also have to lie in $K_P$.
Therefore the concurrence of an element $X \in K_P$ with respect to a positive operator $\Phi: {\cal H}(d_1) \to {\cal H}(d_2)$
equals the concurrence of $X$ with respect to the restriction of $\Phi$ to $U_P$. The same holds for the $I$-fidelity.
Thus Theorem \ref{pos_result} is applicable also in this case, but we have first to find a function on $U_P$
which can serve as determinant in the sense of Definition \ref{Jordandef}. It suffices that this function
is a quadratic form with signature $(+---)$, is positive on $\interior K_P$ and zero on $\partial K_P$.
The second symmetric function $\sigma_2^{d_1}$ on ${\cal H}(d_1)$
has signature $(+-\cdots-)$, is zero on all extremal elements of $H_+(d_1)$ and positive
on all other elements of this cone. Hence its restriction to $U_P$ fulfills the necessary requirements.
We can deduce the following result.

{\theorem \label{pos_general} Let $\Phi: {\cal H}(d_1) \to {\cal H}(d_2)$ be a positive operator and let $X \in H_+(d_1)$ be a matrix of rank not exceeding 2.
Let further $P \subset {\mathbb C}^{d_1}$ be a linear complex subspace of dimension 2 such that the range of $X$ is contained in $P$.
Denote by $U_P$ the subspace of all matrices in ${\cal H}(d_1)$ whose range is contained in $P$.

Then the generalized eigenvalues of the pencil $\sigma_2^{d_2} \circ \Phi|_{U_P} - \lambda \sigma_2^{d_1}|_{U_P}$ are all real.
Denote them by $\lambda_1,\lambda_2,\lambda_3,\lambda_4$ in decreasing order.
Then the concurrence and the $I$-fidelity of $X$ with respect to $\Phi$ are given by
\[ C(\Phi;X) = 2\sqrt{ \sigma_2^{d_2} (\Phi(X)) - \lambda_2 \sigma_2^{d_1} (X) }, \qquad F(\Phi;X) = 2\sqrt{ \sigma_2^{d_2} (\Phi(X)) - \lambda_4 \sigma_2^{d_1} (X) }.
\]
The function $C(\Phi;\cdot)$ (respectively, $F(\Phi;\cdot)$) is linear along any affine subspace which is parallel to
the eigenspace $V_2$ (respectively, $V_4$) corresponding to the eigenvalue $\lambda_2$
(respectively, $\lambda_4$) of the pencil
$\sigma_2^{d_2} \circ \Phi|_{U_P} - \lambda \sigma_2^{d_1}|_{U_P}$.
If $X$ has rank two, then
there exists a matrix in $V_2$ (respectively, $V_4$) which is linearly independent of $X$. This allows to obtain an
optimal decomposition of $X$ as convex combination $X = \mu Y + (1-\mu) Z$ of two rank 1 matrices
$Y,Z \in H_+(d_1)$
such that $C(\Phi;X) = \mu C(\Phi;Y) + (1-\mu) C(\Phi;Z)$ (respectively, $F(\Phi;X) = \mu F(\Phi;Y) + (1-\mu) F(\Phi;Z)$).
\qed }

\smallskip

For a $d_1 \otimes d_2$ bipartite matrix $X$ the concurrence is defined as concurrence of $X$ with respect to the partial trace.
Setting $\Phi$ to the partial trace in the previous theorem, we obtain a formula for the concurrence of bipartite matrices.

{\corollary \label{result_bipartite} Let $X$ be a $d_1 \otimes d_2$ bipartite matrix of rank not exceeding 2.
Let further $P \subset {\mathbb C}^{d_1d_2}$ be a linear complex subspace of dimension 2 such that the range of $X$ is contained in $P$.
Denote by $U_P$ the subspace of all matrices in ${\cal H}(d_1d_2)$ whose range is contained in $P$.
Define the two quadratic forms $Q_1: A \mapsto 2((tr\,A)^2 - tr((tr_1A)^2))$ and
$Q_2: A \mapsto (tr\,A)^2 - tr\,A^2$ on ${\cal H}(d_1d_2)$.

Then the generalized eigenvalues of the pencil $Q_1|_{U_P} - \lambda Q_2|_{U_P}$ are all real.
Denote them by $\lambda_1,\lambda_2,\lambda_3,\lambda_4$ in decreasing order.
Then the concurrence and the $I$-fidelity of $X$ are given by
\[ C(X) = \sqrt{ Q_1(X) - \lambda_2 Q_2(X) }, \qquad F(X) = \sqrt{ Q_1(X) - \lambda_4 Q_2(X) }.
\]
The function $C(\Phi;\cdot)$ (respectively, $F(\Phi;\cdot)$) is linear along any affine subspace which is parallel to
the eigenspace $V_2$ (respectively, $V_4$) corresponding to the eigenvalue $\lambda_2$
(respectively, $\lambda_4$) of the pencil
$Q_1|_{U_P} - \lambda Q_2|_{U_P}$.
If $X$ has rank two, then
there exists a matrix in $V_2$ (respectively, $V_4$) which is linearly independent of $X$. This allows to obtain an
optimal decomposition of $X$ as convex combination $X = \mu Y + (1-\mu) Z$ of two rank 1 matrices
$Y,Z \in H_+(d_1d_2)$
such that $C(X) = \mu C(Y) + (1-\mu) C(Z)$ (respectively, $F(X) = \mu F(Y) + (1-\mu) F(Z)$).
\qed }

\smallskip

The quadratic form $Q_1$ in the corollary can be replaced by the form $A \mapsto 2((tr\,A)^2 - tr((tr_2A)^2))$
or by ${\cal S}_{d_1} \otimes {\cal S}_{d_2}$, where ${\cal S}_d$ is the universal inverter
on ${\cal H}(d)$ as defined in \cite{Rungta01}. All these forms are equal on the pure states and hence on the
boundary of $K_P$.

{\remark While the form $A \mapsto 2((tr\,A)^2 - tr((tr_2A)^2))$ is generated by the determinant of the Lorentz-positive map $tr_2$
and hence fulfills condition (\ref{main_cond}), the argument for ${\cal S}_{d_1} \otimes {\cal S}_{d_2}$,
or for any other quadratic form that coincides with $Q_1$ on the boundary of $K_P$, is as follows.
Let $J$ be a regular quadratic form on ${\mathbb R}^n$ with signature $(+-\cdots-)$. Let further $Q,Q'$ be
quadratic forms fulfilling $x^TQx = x^TQ'x$ for all $x$ such that $x^TJx = 0$. Then there exists a real number
$\lambda$ such that $Q' = Q + \lambda J$. Therefore $Q - \lambda J$ and $Q' - \lambda J$ are essentially
the same pencils.}

\section{Density matrices of 2-edge graphs}

In this section we apply the results of the previous section to density matrices of graphs with 2 edges.
Let $G$ be a graph and $M(G)$ its adjacency matrix, i.e.\ a symmetric matrix of size $n \times n$, where $n$
is the number of vertices of $G$, its $(k,l)$ element given by 1 if the vertices $k,l$ are connected
by an edge and by 0 otherwise. Let further $\Delta(G)$ be the degree matrix of the graph, i.e.\ the unique
diagonal matrix such that the combinatorial Laplacian $L(G) = \Delta(G) - M(G)$ contains the vector
$(1,1,\dots,1)^T$ in its kernel. The density matrix $\rho_G$ of $G$ is defined as $\frac{L(G)}{tr\,L(G)} = \frac{L(G)}{2n_e}$,
where $n_e$ is the number of edges.
For details see \cite{Braunstein06}. Note that $n_e$ is an upper bound for the rank of the density matrix.

If the vertices of the graph are arranged in an array of size $d_1 \times d_2$, then its density matrix inherits
a natural $d_1 \otimes d_2$ bipartite structure.
In \cite{HilManSev06} the formula of Wootters \cite{Wootters98} was used to compute the concurrences of all
graphs on 4 vertices, arranged in an $2 \times 2$ array. It turned out that either the density matrix was separable
or the concurrence was equal to $\frac{1}{n_e}$. The corresponding optimal
decomposition of the nonseparable density matrices derives from the decomposition of the combinatorial Laplacian in
the combinatorial Laplacians of all subgraphs with one edge.

In this section we compute the concurrences and $I$-fidelities of all density matrices of graphs having rank 2.
These are the density matrices of all graphs with 2 edges and of all graphs with 3 edges arranged in a closed loop.
For the computation we used the universal inverter as defined in \cite{Rungta01}. Suppose the vertices of the
graph $G$ in question are arranged in an array of size $d_1 \times d_2$, giving rise to a $d_1 \otimes d_2$ bipartitioned
density matrix $\rho_G$. Let the form $Q_1$ on ${\cal H}(d_1d_2)$ from Corollary \ref{result_bipartite} be defined by ${\cal S}_{d_1} \otimes {\cal S}_{d_2}$,
i.e.\ acting as $Q_1: \rho_G \mapsto \langle \rho_G, tr\,\rho_G I_{d_1d_2} - I_{d_1} \otimes tr_1\rho_G - tr_2\rho_G \otimes
I_{d_2} + \rho_G \rangle$, and let the form $Q_2$ be defined by $Q_2: \rho_G \mapsto (tr\,\rho_G)^2 - tr\,\rho_G^2$.
Let further $U$ be the 4-dimensional subspace of ${\cal H}(d_1d_2)$ consisting of all matrices that have the same range as $\rho_G$.
Then the concurrence and the $I$-fidelity of $\rho_G$ are given by
$C(\rho_G) = \sqrt{ Q_1(\rho_G) - \lambda_2 Q_2(\rho_G) }$, $F(\rho_G) = \sqrt{ Q_1(\rho_G) - \lambda_4 Q_2(\rho_G) }$,
where $\lambda_1,\lambda_2,\lambda_3,\lambda_4$ are the generalized eigenvalues of the pencil $Q_1|_U - \lambda Q_2|_U$
in decreasing order.

Below we depict the graphs, each representing one isomorphism class, and list the values of the corresponding
concurrences and $I$-fidelities, the generalized eigenvalues of the pencil $Q_1|_U - \lambda Q_2|_U$ and the values
of the quadratic forms $Q_1,Q_2$ on $\rho_G$.

\begin{center}
\vskip 0.5cm
\begin{tabular}[c]{ccccccccc}
\includegraphics[height=0.45in,width=0.45in]{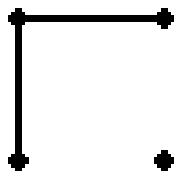}
& \ &
\includegraphics[height=0.45in,width=0.45in]{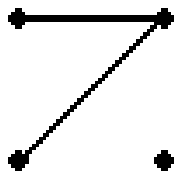}
& \ &
\includegraphics[height=0.45in,width=0.45in]{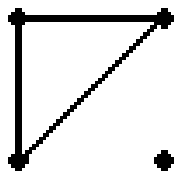}
& \ &
\includegraphics[height=0.45in,width=0.45in]{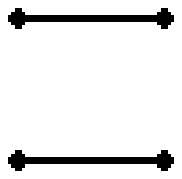}
& \ &
\includegraphics[height=0.45in,width=0.45in]{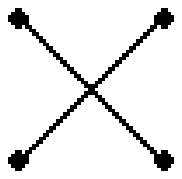}\\
Ia &  & Ib &  & Ic &  & II & & III
\end{tabular}
\vskip 1cm
\begin{tabular}[c]{c}
\begin{tabular}[c]{ccccccc}
\includegraphics[height=0.45in,width=0.75in]{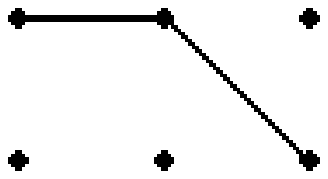}
& \ &
\includegraphics[height=0.45in,width=0.75in]{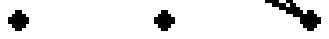}
& \ &
\includegraphics[height=0.45in,width=0.75in]{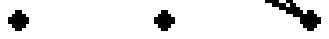}
& \ &
\includegraphics[height=0.45in,width=0.75in]{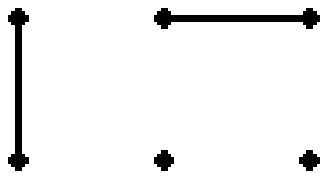}\\
IVa & & IVb & & IVc & & V
\end{tabular}
\\
\begin{tabular}[c]{ccccccc}
\includegraphics[height=0.45in,width=0.75in]{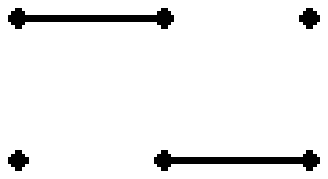}
& \ &
\includegraphics[height=0.45in,width=0.75in]{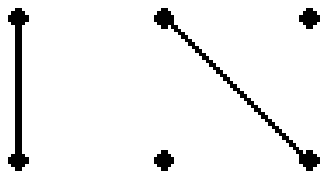}
& \ &
\includegraphics[height=0.45in,width=0.75in]{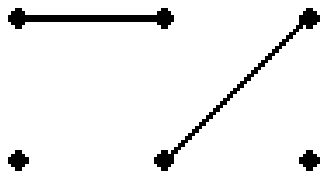}
& \ &
\includegraphics[height=0.45in,width=0.75in]{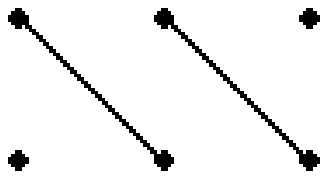}\\
VI & & VII && VIII & & IX
\end{tabular}
\end{tabular}
\vskip 1cm
\begin{tabular}[c]{ccccc}
\includegraphics[height=0.45in,width=1.05in]{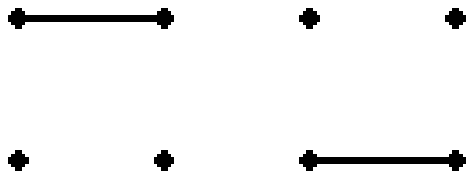}
& \ &
\includegraphics[height=0.45in,width=1.05in]{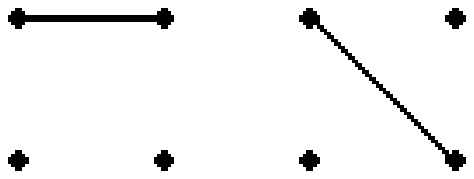}
& \ &
\includegraphics[height=0.45in,width=1.05in]{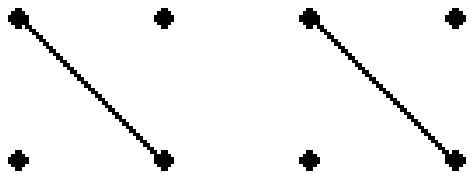}\\
X &  & XI &  & XII
\end{tabular}
\vskip 1cm
\begin{tabular}[c]{c}
\begin{tabular}[c]{ccccccc}
\includegraphics[height=0.75in,width=0.75in]{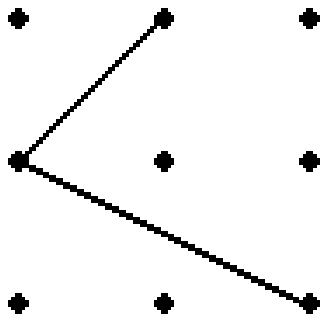}
& \ &
\includegraphics[height=0.75in,width=0.75in]{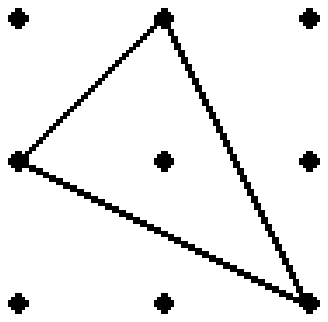}
& \ &
\includegraphics[height=0.75in,width=0.75in]{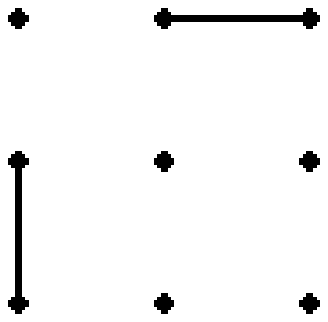}\\
XIIIa &  & XIIIb &  & XIV
\end{tabular}
\\
\begin{tabular}[c]{ccccc}
\includegraphics[height=0.75in,width=0.75in]{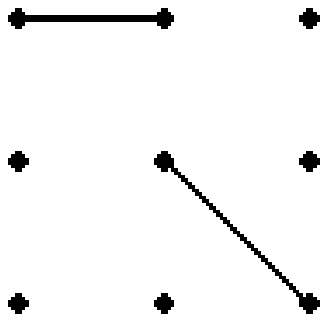}
& \ &
\includegraphics[height=0.75in,width=0.75in]{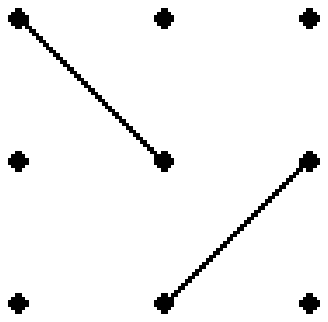}
& \ &
\includegraphics[height=0.75in,width=0.75in]{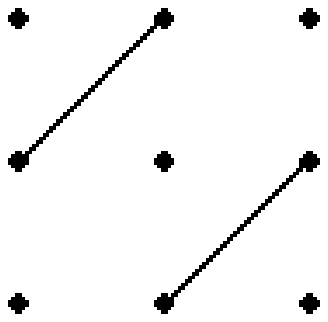}\\
XV & & XVI & & XVII
\end{tabular}
\end{tabular}
\vskip 1cm
\begin{tabular}[c]{ccccc}
\includegraphics[height=0.75in,width=1.05in]{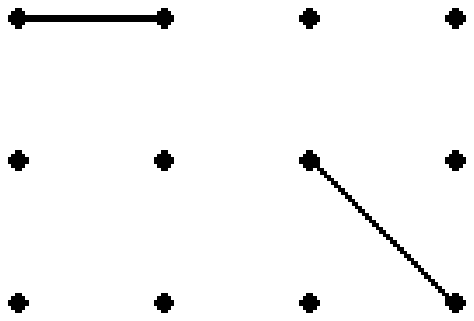}
& \ &
\includegraphics[height=0.75in,width=1.05in]{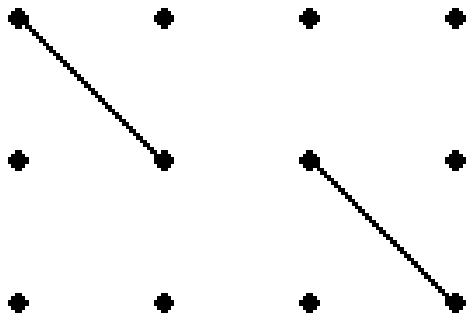}
& \ &
\includegraphics[height=1.05in,width=1.05in]{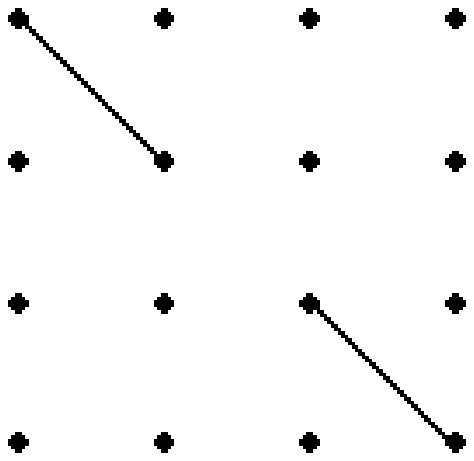}\\
XVIII & & XIX & & XX
\end{tabular}
\end{center}

\bigskip

\begin{center}
\begin{tabular}{c||c|c|c|c||c|c||c|c}
Type & $\lambda_1$ & $\lambda_2$ & $\lambda_3$ & $\lambda_4$ & $Q_1(\rho_G)$ & $Q_2(\rho_G)$ & $C(\rho_G)$ & $F(\rho_G)$ \\
\hline\hline
Ia & $1/3$ & $1/3$ & $-1/3$ & $-1/3$ & $1/8$ & $3/8$ & $0$ & $1/2$ \\
Ib & & & & & $3/8$ & $3/8$ & $1/2$ & $\sqrt{2}/2$ \\
Ic & & & & & $5/18$ & $1/2$ & $1/3$ & $2/3$ \\
\hline
II & $0$ & $0$ & $0$ & $0$ & $0$ & $1/2$ & $0$ & $0$ \\
\hline
III & $1$ & $1$ & $-1$ & $-1$ & $1/2$ & $1/2$ & $0$ & $1$ \\
\hline\hline
IVa & $2/3$ & $2/3$ & $-2/3$ & $-2/3$ & $1/2$ & $3/8$ & $1/2$ & $\sqrt{3}/2$ \\
IVb & & & & & $3/4$ & $3/8$ & $\sqrt{2}/2$ & $1$ \\
IVc & & & & & $5/9$ & $1/2$ & $\sqrt{2}/3$ & $2\sqrt{2}/3$ \\
\hline
V & $1/2$ & $1/2$ & $-1/2$ & $-1/2$ & $1/4$ & $1/2$ & $0$ & $\sqrt{2}/2$ \\
\hline
VI & $3/4$ & $3/4$ & $-3/4$ & $-3/4$ & $3/8$ & $1/2$ & $0$ & $\sqrt{3}/2$ \\
\hline
VII & $1/2$ & $1/2$ & $-1/2$ & $-1/2$ & $1/2$ & $1/2$ & $1/2$ & $\sqrt{3}/2$ \\
\hline
VIII & $1/4$ & $1/4$ & $-1/4$ & $-1/4$ & $3/8$ & $1/2$ & $1/2$ & $\sqrt{2}/2$ \\
\hline
IX & $5/4$ & $-1/4$ & $-1/4$ & $-3/4$ & $5/8$ & $1/2$ & $\sqrt{3}/2$ & $1$ \\
\hline\hline
X & $1$ & $1$ & $-1$ & $-1$ & $1/2$ & $1/2$ & $0$ & $1$ \\
\hline
XI & $1/2$ & $1/2$ & $-1/2$ & $-1/2$ & $1/2$ & $1/2$ & $1/2$ & $\sqrt{3}/2$ \\
\hline
XII & $3/2$ & $-1/2$ & $-1/2$ & $-1/2$ & $3/4$ & $1/2$ & $1$ & $1$ \\
\hline\hline
XIIIa & $5/3$ & $-1/3$ & $-1/3$ & $-1$ & $7/8$ & $3/8$ & $1$ & $\sqrt{5}/2$ \\
XIIIb & & & & & $5/6$ & $1/2$ & $1$ & $2\sqrt{3}/3$ \\
\hline
XIV & $1$ & $1$ & $-1$ & $-1$ & $1/2$ & $1/2$ & $0$ & $1$ \\
\hline
XV & $3/4$ & $3/4$ & $-3/4$ & $-3/4$ & $5/8$ & $1/2$ & $1/2$ & $1$ \\
\hline
XVI & $3/2$ & $-1/2$ & $-1/2$ & $-1/2$ & $3/4$ & $1/2$ & $1$ & $1$ \\
\hline
XVII & $3/2$ & $-1/2$ & $-1/2$ & $-1/2$ & $3/4$ & $1/2$ & $1$ & $1$ \\
\hline\hline
XVIII & $1$ & $1$ & $-1$ & $-1$ & $3/4$ & $1/2$ & $1/2$ & $\sqrt{5}/2$ \\
\hline
XIX & $7/4$ & $-1/4$ & $-3/4$ & $-3/4$ & $7/8$ & $1/2$ & $1$ & $\sqrt{5}/2$ \\
\hline\hline
XX & $2$ & $0$ & $-1$ & $-1$ & $1$ & $1/2$ & $1$ & $\sqrt{6}/2$
\end{tabular}
\end{center}

\bigskip

The graphs of subtypes distinguished by letters lead to density matrices which have the same range
and hence share the space $U$ and the spectrum $\lambda_1,\dots,\lambda_4$ of the corresponding pencil.

\section{Conclusions}

The concurrence of positive operators or bipartite positive semidefinite matrices is defined as the convex roof
of the square root of a certain quadratic function defined on the positive semidefinite rank 1 matrices in the input space ${\cal H}(d)$,
the space of complex hermitian matrices of size $d \times d$. The convex roof is the largest convex extension of the
function in question to the convex hull of the rank 1 matrices, i.e.\ to the cone of positive semidefinite matrices in
${\cal H}(d)$. Our ability to compute concurrences hence depends on our ability to compute such convex roofs.

The concept of convex roof can be generalized to other regular convex cones than the positive semidefinite cone. It can be defined
as the largest convex extension of a function defined on the extremal elements of the cone to the cone itself. In this paper we
derived an explicit expression for the convex roof of a certain class of functions and a certain class of cones,
namely cones that are generated by quadratic forms with signature $(+-\cdots-)$, i.e.\ linear images of the Lorentz cone.
For the same class we are also able to compute the concave roof. This result is the main technical contribution of the present
paper and is formalized in Theorem \ref{main}.

It allows us to obtain explicit formulae for the concurrence of Lorentz-positive maps as defined in Definition \ref{conc_lor},
as well as for a function called $I$-fidelity, which is defined via the concave roof in Definition \ref{fid_lor}.
These formulae are provided in Theorems \ref{concl2} and \ref{fidlm}. From Lemma \ref{equal_functions} it follows that
if the input space of a positive operator is ${\cal H}(2)$, then computing its concurrence and $I$-fidelity can be reduced to
computing the concurrence or $I$-fidelity of a Lorentz-positive map. This allows us to obtain explicit formulae for these
quantities, provided in Theorem \ref{pos_result}. More generally, it allows us to compute the concurrence and $I$-fidelity
of any matrix of rank not exceeding two with respect to any positive operator. These formulae are provided by
Theorem \ref{pos_general}. Further we investigate the relation between the concurrence of positive operators
and bipartite matrices. In Lemma \ref{pos_bip} we show that the notions of concurrence and $I$-fidelity
for a bipartite matrix are essentially equivalent to those for a completely positive operator.
Hence we are also able to compute the concurrence and $I$-fidelity of rank two bipartite matrices. The
corresponding formulae are provided in Corollary \ref{result_bipartite}. In all cases the optimal decomposition
yielding the value of the concurrence or $I$-fidelity contains two pure states. The optimal decomposition can be obtained
via the eigenvector of the corresponding matrix pencil to the second largest generalized eigenvalue for the concurrence
and the smallest one for the $I$-fidelity.

\bibliography{quantinf,convexity}
\bibliographystyle{plain}

\end{document}